\documentclass[twocolumn,showpacs,preprintnumbers,amsmath,amssymb]{revtex4}

\usepackage{graphicx}
\usepackage{dcolumn}
\usepackage{bm}

\newcommand{\be}{\begin{equation}} 
\newcommand{\ee}{\end{equation}} 
\newcommand{\bea}{\begin{eqnarray}} 
\newcommand{\eea}{\end{eqnarray}}  
\newcommand{\bean}{\begin{eqnarray*}} 
\newcommand{\eean}{\end{eqnarray*}}

\def\lsim{\raise 0.4ex\hbox{$<$}\kern -0.8em\lower 0.62ex\hbox{$\sim$}} 
\def\gsim{\raise 0.4ex\hbox{$>$}\kern -0.7em\lower 0.62ex\hbox{$\sim$}} 
\newcommand{\bk}{{\bf k}}

\begin{document}


\title{Gravitational evolution of a perturbed lattice and its fluid limit}   
\author{M. Joyce}   
\affiliation{Laboratoire de Physique Nucl\'eaire et de Hautes Energies,  
 Universit\'e de Paris VI, 4, Place Jussieu, 
Tour 33 -RdC, 75252 Paris Cedex 05, France.} 
\author{B. Marcos}   
\affiliation{Laboratoire de Physique Th\'eorique,  
         Universit\'e de Paris XI, B\^atiment 210,  
        91405 Orsay, France.}  
\author{A. Gabrielli}   
\affiliation{ SMC-INFM \& ISC-CNR,
Dipartimento di Fisica, 
Universit\`a ``La Sapienza'',
P.le A. Moro 2,
I-00185 Rome,
Italy.} 
\author{T. Baertschiger}   
\affiliation{ Dipartimento di Fisica, 
Universit\`a ``La Sapienza'',
P.le A. Moro 2,
I-00185 Rome,
Italy.}
\author{F. Sylos Labini}   
\affiliation{ ``E. Fermi'' Center, Via Panisperna 89 A, Compendio del 
Viminale, 00184 - Rome, Italy.}     


\begin{abstract}
We apply a simple linearization, well known in solid state physics,
to approximate the evolution at early times of cosmological $N$-body 
simulations of gravity. In the limit that the initial perturbations,
applied to an infinite perfect lattice, are at wavelengths much
greater than the lattice spacing $l$ the evolution is exactly 
that of a pressureless self-gravitating fluid treated in  
the analagous (Lagrangian) linearization, with the Zeldovich 
approximation as a sub-class of asymptotic solutions.
Our less restricted approximation allows one to trace the evolution of the 
{\it discrete} distribution until the time when particles 
approach one another (i.e. ``shell crossing''). We calculate 
modifications of the fluid evolution, explicitly dependent 
on $l$ i.e. discreteness effects in the N body simulations.
We note that these effects become increasingly important 
as the initial red-shift
is increased at fixed $l$. The possible advantages of using a body
centred cubic, rather than simple cubic, lattice are pointed out.
\end{abstract}

\maketitle
In current cosmological theories the physics of
structure formation in the universe reduces, over 
a large range of scales, to 
understanding the evolution of clustering 
under Newtonian gravity, with only a simple 
modification of the dynamical 
equations due to the expansion of the Universe.
The primary
instrument for solving this problem is numerical
$N$-body simulation (NBS, see e.g.\cite{nbs-standard-references}). 
These simulations are most usually 
started from configurations which are simple cubic (sc) lattices 
perturbed in a manner prescribed by a
theoretical cosmological model. In this letter we observe that, up
to a change in sign in the force, the 
initial configuration is identical to the Coulomb lattice
(or Wigner crystal) in solid state physics 
(see e.g. \cite{pines}), and we exploit this analogy to develop 
an approximation to the evolution of these simulations.
We show that one obtains, for long wavelength perturbations,
the evolution predicted by an analagous fluid description
of the self-gravitating system, and in
particular, as a special case, the Zeldovich 
approximation \cite{zeldovich-ZApaper}. Further 
we can study precisely the deviations from this fluid-like behaviour at 
shorter wavelengths arising from the discrete nature of the 
system. This analysis should be a useful step towards a precise 
quantitative understanding, which is currently lacking, of the 
role of discreteness in cosmological NBS 
(see e.g. \cite{discreteness-references, discreteness-tbmjfsl, 
discreteness-hamana}).
One simple conclusion, for example, is that a body centred 
cubic (bcc) lattice may be a better choice of discretisation,
as its spectrum has only growing modes with exponents
bounded above by that of fluid linear theory.

The equation of motion of particles moving under their
mutual self-gravity is \cite{nbs-standard-references}
\be
{\ddot {\bf x} }_i +
2 H (t) {\dot {\bf x} }_i
= -\frac{1}{a^3} \sum_{i\neq j} 
\frac{G m_j ({\bf x}_{i}-{\bf x}_j) }{|{\bf x}_{i}-{\bf x}_{j}|^3}\,.
\label{eom}
\ee
Here dots denote derivatives with respect to time $t$,
${\bf x}_i$ is the comoving position of the $i$th particle,
of mass $m_i$, related to the physical coordinate by 
${\bf r}_i=a(t) {\bf x}_i$, where $a(t)$ is the scale factor 
of the background cosmology with Hubble constant
$H(t)=\frac{\dot a}{a}$.
We treat a system of $N$ point particles, of equal 
mass $m$, initially placed on a Bravais lattice, with 
periodic boundary conditions.
Perturbations from the Coulomb lattice are described 
simply by  Eq.~(\ref{eom}), with $a(t)=1$ and $Gm^2 \rightarrow -e^2$
(where $e$ is the electronic charge).
As written in Eq.~(\ref{eom}) the infinite sum giving the
force on a particle is not explicitly well defined. It is 
calculated by solving the Poisson equation for the potential,
with the mean mass density subtracted in the
source term. In the cosmological case this is appropriate
as the effect of the mean density is absorbed in the 
Hubble expansion; in the case of the Coulomb lattice 
it corresponds to the assumed presence of an oppositely charged
neutralising background.

We consider now perturbations about the perfect lattice.
It is convenient to adopt the notation 
${\bf x}_i(t)={\bf R} + {\bf u}({\bf R},t)$
where ${\bf R}$ is the lattice vector of the $i$th particle
(which we can consider as its Lagrangian coordinate),
and ${\bf u}({\bf R},t)$ is the displacement of the particle from 
{\bf R}. Expanding to linear order in ${\bf u}({\bf R},t)$
about the equilibrium lattice configuration (in which the force on
each particle is exactly zero), we obtain
\be
{\bf {\ddot u}}({\bf R},t) 
+2 H {\bf {\dot u}}({\bf R},t) 
= -\frac{1}{a^3} \sum_{{\bf R}'} 
{\cal D} ({\bf R}- {\bf R}') {\bf u}({\bf R}',t)\,. 
\label{linearised-eom}
\ee
The matrix ${\cal D}$ is known in solid
state physics, for any interaction, as the
{\it dynamical matrix} (see e.g. \cite{pines}).  
For gravity we have 
${\cal D}_{\mu \nu} ({\bf R} \neq {\bf 0})=
Gm(\frac{\delta_{\mu \nu}}{R^3}
-3\frac{R_\mu R_\nu}{R^5})$
(where $\delta_{\mu \nu}$ is the Kronecker delta),
and ${\cal D}_{\mu \nu} ({\bf 0})= 
-\sum_{{\bf R} \neq {\bf 0}} {\cal D}_{\mu \nu} ({\bf R})$
\footnote{For conciseness of notation we have left 
implicit in these expressions the sum over the copies 
which results from the periodic boundary conditions.}.
 
From the Bloch theorem for lattices 
it follows that ${\cal D}$ is diagonalised by plane waves in 
reciprocal space. Defining the Fourier transform by 
${\bf {\tilde u}}({\bf k},t)= \sum_{{\bf R}} e^{-i {\bf k}\cdot{\bf R}}
{\bf u}({\bf R},t)$ and its inverse as 
${\bf u}({\bf R},t)= \frac{1}{N} \sum_{{\bf k}} e^{i {\bf k}\cdot{\bf R}}
 {\bf {\tilde u}}({\bf k},t)$ (where the sum is over the first 
Brillouin zone), Eq.~(\ref{linearised-eom}) gives
\be
{\bf \ddot{{\tilde u}}} ({\bf k},t) 
+ 2 H (t) {\bf \dot{{\tilde u}}} ({\bf k},t) 
= -\frac{1}{a^3} {\cal {\tilde D}} ({\bf k}) {{\bf {\tilde u}}}({\bf k},t) 
\ee
where ${\cal {\tilde D}} ({\bf k})$, the Fourier transform (FT) of 
${\cal D} ({\bf R})$, 
is a symmetric $3 \times 3$ matrix for each ${\bf k}$. 
Diagonalising it one can determine, for each ${\bf k}$,  
three orthonormal eigenvectors ${\bf e}_n ({\bf k})$ and their
eigenvalues $\omega_n^2({\bf k})$ ($n=1,2,3$), which
obey \cite{pines} the Kohn sum rule 
$\sum_n \omega_n^2({\bf k}) = -4 \pi G \rho_0$,
where $\rho_0$ is the mean mass density.

Given the initial displacements and velocities 
at a time $t=t_0$, the dynamical evolution is then 
given as 
\bea
{\bf u}({\bf R}, t) &=\frac{1}{N} 
\sum_{{\bf k}} \sum_{n}
[{\bf {\tilde u}}({\bf k},t_0)\cdot{\bf {\hat e}}_n({\bf k}) 
U_n({\bf k},t)
\nonumber \\
&+ {\bf \dot{\tilde u}}({\bf k},t_0)\cdot{\bf {\hat e}}_n({\bf k})  
V_n({\bf k},t) ] {\bf {\hat e}}_n({\bf k})e^{i {\bf k}\cdot{\bf R}}
\label{linearised-evolution-general}
\eea
where $U_n({\bf k},t)$ and $V_n({\bf k},t)$ are a set of linearly 
independent solutions of the mode equations
\be
{\ddot{f}} + 2 H {\dot{f}}= -\frac{\omega_n^2({\bf k})}{a^3} f
\label{mode-equation}
\ee
chosen so that
$U_n({\bf k},t_0)=1$, $\dot{U}_n({\bf k},t_0)=0$, 
$V_n({\bf k},t_0)=0$, $\dot{V}_n({\bf k},t_0)=1$. 

\begin{figure}
\resizebox{8cm}{!}{\includegraphics*{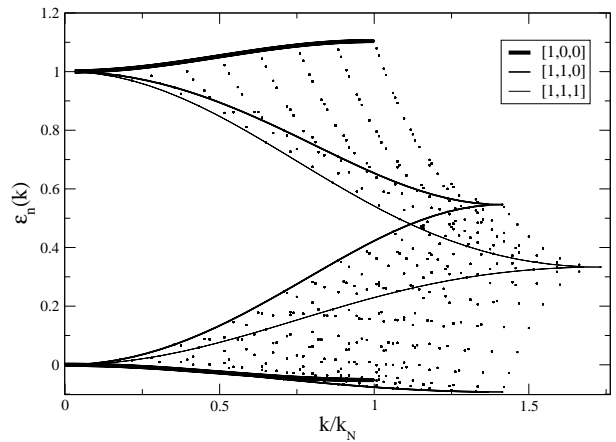}}
\caption{Eigenvalues  $\epsilon_n({\bf k})$ for a sc lattice. 
The lines connect eigenvectors with ${\bf k}$ 
in the specific directions indicated. Note that the 
two acoustic branches are degenerate in the $[1,0,0]$ 
and $[1,1,1]$ directions.
\label{fig1}}
\end{figure}

In Fig.~\ref{fig1} are shown the eigenvalues of the
dynamical matrix, for gravity, for a $16^3$ sc 
lattice, determined numerically by applying the 
linearization to a standard Ewald 
summation of the gravitational force (see e.g. \cite{ewald}).
For convenience the eigenvalues  have been normalized,
with $\epsilon_n({\bf k})=-\frac{\omega_n^2({\bf k})}{4 \pi G \rho_0}$,
and they are plotted, as a function of the modulus $k\equiv|{\bf k}|$, 
normalized to the Nyquist frequency $k_N=\pi/l$, where $l$ 
is the lattice spacing. This diagonalisation can be 
performed rapidly even for the largest 
lattices used in current cosmological NBS, but
the figure remains essentially unchanged except for an
increase in the density of the eigenvalues.
The lines in the figure connect the eigenvectors
along some specific chosen directions, making the
characteristic branch structure of the eigenvectors
evident. It can be shown 
\cite{pines} that ${\cal D}_{\mu\nu}({\bf k} \rightarrow 0)=  
- {\hat k}_\mu {\hat k}_\nu 4 \pi G \rho_0$ (where 
${\bf \hat k}={\bf k}/k$), so the branch with eigenvalue 
tending to $-4 \pi G \rho_0$ is longitudinal (in this limit). 
In the Coulomb lattice this is the {\it optical} branch, describing 
oscillations for $k \rightarrow 0$ with plasma frequency 
$\omega_p^2=4 \pi e^2 n_0/m$ 
(where $n_0$ is the electronic number density). There are 
then also two {\it acoustic} branches
with eigenvalues tending to zero as $k \rightarrow 0$  and which 
become purely transverse in this limit. A striking feature 
of Fig.\ref{fig1} is 
that there are eigenvectors with $\epsilon_n({\bf k}) < 0$, which
correspond to unstable modes for the Coulomb system, with solutions to 
Eq.~(\ref{mode-equation}) 
$U_n({\bf k},t)= \cosh(|\omega_n({\bf k})|t)$ and
$V_n({\bf k},t)=(1/|\omega_n({\bf k})|) \sinh(|\omega_n({\bf k})|t)$
(taking  $a=1$ and $t_0=0$). Thus the sc Coulomb lattice is 
unstable to perturbations, which is not an unexpected result:
the ground state of this classical system is known to be 
the bcc lattice \cite{wc-ground state}, and these unstable modes
in the sc lattice correspond to
instabilities towards such lower energy configurations. For the case
of gravity, in a static universe, these modes are
sinusoidally oscillating, while the modes 
$\epsilon_n({\bf k}) > 0$ describe the expected
exponential instabilities. Note further that, since the Kohn sum rule 
can be written $\sum_n \epsilon_n({\bf k})= 1$,  the appearance
of modes with $\epsilon_n({\bf k})>1$ is only possible when
there are modes with $\epsilon_n({\bf k}) < 0$. Without 
calculation we can thus conclude that a bcc lattice
will have only unstable modes in the case of gravity,
and that $\epsilon_n({\bf k}) \leq 1$. We will return to this
point below.

The damping term coming from the expansion of the universe
modifies these solutions to Eq.~(\ref{mode-equation}). 
For the case of an Einstein de Sitter (EdS, flat
matter dominated) universe, for which  $H^2(t)=
\frac{8\pi G\rho_0}{3a^3}$ and thus $a=(t/t_0)^{2/3}$, 
we find
\bea
U_n({\bf k},t)&=&\frac{
\alpha_n^+({\bf k}) (t/t_0)^{\alpha_n^-({\bf k})}
+
\alpha_n^-({\bf k}) (t/t_0)^{-\alpha_n^+({\bf k})}}  
{\alpha_n^+({\bf k}) + \alpha_n^-({\bf k})} 
\nonumber
\\ 
V_n({\bf k},t)&=& t_0\frac{
(t/t_0)^{\alpha_n^-({\bf k})}
-(t/t_0)^{-\alpha_n^+({\bf k})}}
{\alpha_n^+({\bf k}) + \alpha_n^-({\bf k})}
\label{Un+Vn-EdS}
\eea
where 
$\alpha_n^\pm({\bf k})=\frac{1}{6}[\sqrt{1+24 \epsilon_n({\bf k})} \pm 1]$.
Thus for $\epsilon_n({\bf k})>0$ there are, as in the
static case, both a growing and a decaying solution. For
$\epsilon_n({\bf k})< 0$ the solutions 
are all power-law decaying. For 
$\epsilon_n({\bf k}) < -\frac{1}{24}$, there is 
a weak remnant of the static universe oscillating 
behaviour: $\alpha_n^\pm ({\bf k})$ are then 
complex, and it is simple to show that the mode
functions are a product of a power 
law $(t/t_0)^{-\frac{1}{6}}$ and a sinusoidal
oscillation periodic in the logarithm of the 
evolution time $\ln (t/t_0)$.

Let us now consider the case that the initial 
fluctuations contain only modes s.t. $kl \ll 1$.
We have then simply for each ${\bf k}$ 
the longitudinal mode ${\bf e}_1({\bf k})={\hat{\bf k}}$,
with $\epsilon_1({\bf k})=1$, 
and two transverse  modes with zero eigenvalues. 
Using the corresponding mode functions from  
Eq.~(\ref{Un+Vn-EdS}), and Eq.~(\ref{linearised-evolution-general}),
a simple calculation shows that 
\bea
{\bf u}({\bf R},t) &= {\bf u}_{\perp} ({\bf R},t_0)
+ {\bf u}_{\parallel}({\bf R},t_0)
\left[\frac{3}{5} (\frac{t}{t_0})^{\frac{2}{3}}
+\frac{2}{5} (\frac{t}{t_0})^{-1}\right]
\nonumber\\
&+{\bf v}_{\parallel}({\bf R},t_0) t_0
\left[\frac{3}{5} (\frac{t}{t_0})^{\frac{2}{3}}
-\frac{3}{5} (\frac{t}{t_0})^{-1}\right]
\nonumber\\
&+ {\bf v}_{\perp}({\bf R},t_0) 3t_0
\left[1-(\frac{t}{t_0})^{-\frac{1}{3}}\right]
\label{displacement-fluid-limit}
\eea
where we have decomposed the particle displacements 
and peculiar velocities 
(${\bf v}({\bf R},t) \equiv 
{\dot {\bf r}}_i - H{\bf r}_i 
=a{\bf {\dot u}}({\bf R},t)$)
into an irrotational (curl-free)
part 
${\bf a}_{\parallel} ({\bf R})= \frac{1}{N} 
\sum_{{\bf k}} 
({\bf a}({\bf R}) \cdot{\bf {\hat k}}){\bf {\hat k}} \, 
e^{i {\bf k}.{\bf R}}$,
and a rotational part 
${\bf a}_{\perp}={\bf a} - {\bf a}_{\parallel}$.
Using the definition of the peculiar 
gravitational acceleration 
${\bf g} ({\bf R}, t) \equiv
{\ddot {\bf r}}_i - \frac{\ddot {a}}{a}{\bf r}_i 
= a[{\bf {\ddot u}} + 2 H {\bf {\dot u}}]$, it is simple to show,
using Eq.~(\ref{linearised-eom}), that
${\bf g} ({\bf R}, t_0)=4\pi G {\rho_0} {\bf u}_{\parallel}({\bf R},t_0)=
\frac{2}{3 t_0^2}{\bf u}_{\parallel}({\bf R}, t_0)$.
Using this expression in Eq.~(\ref{displacement-fluid-limit}),
the displacement of each particle with respect to its initial 
position (i.e. ${\bf u}({\bf R},t) -{\bf u}({\bf R},t_0)$)
can be written solely in terms of the initial
gravitational field ${\bf g} ({\bf R}, t_0)$
and the components of the initial peculiar velocity,
${\bf v}_{\perp}({\bf R},t_0)$ and
${\bf v}_{\parallel}({\bf R},t_0)$. 
It is then easy to verify that the solution 
in Eq.~(\ref{displacement-fluid-limit}) corresponds 
exactly to that derived in \cite{buchert1992}, from 
a linearization of the Lagrangian equations for a self-gravitating fluid,
for the displacements of fluid elements with 
respect to their Lagrangian coordinates
\footnote{The Lagrangian coordinate ${\bf X}$ used 
in \cite{buchert1992} is the position of the
particle at $t=t_0$ i.e. 
${\bf X}={\bf R}+ {\bf u} ({\bf R},t_0)$.}. 
As discussed in \cite{buchert1992} there are several
limits of this expression which correspond to
the so called Zeldovich approximation (ZA), which assumes
\cite{zeldovich-ZApaper}  a decomposition of 
${\bf u} ({\bf R}, t)$ into a product
of a function of time and a single vector field 
defined at ${\bf R}$.
The most commonly used form of this approximation  
takes ${\bf u}_\perp({\bf R},t_0)=0={\bf v}_\perp ({\bf R}, t)$ and
${\bf u}_\parallel ({\bf R},t_0)=\frac{3}{2}{\bf v}_\parallel 
({\bf R},t_0)t_0$. This corresponds to setting the coefficients of
all but the growing mode in Eq.~(\ref{displacement-fluid-limit}) 
to zero i.e. it imposes directly the asymptotic behaviour of 
the general solution. We then have simply
${\bf u} ({\bf R}, t) = \frac{3}{2} {\bf g} ({\bf R}, t_0) t_0^2 
(t/t_0)^{\frac{2}{3}}$
which is precisely the solution used standardly in setting up 
initial conditions for cosmological NBS (e.g. \cite{nbs-standard-references}).

This result provides a direct analytical derivation 
explaining precisely the well documented success 
(see e.g. \cite{melott-ZA}) of the ZA in describing 
the evolution of cosmological NBS, in particular 
in  ``truncated'' forms of the approximation in which initial
short wavelength power is filtered \cite{truncatedZA}. 
The eigenvectors and the spectrum of eigenvalues contain,
however, much more than this fluid limit. The
expression Eq.~(\ref{linearised-evolution-general}) 
gives an approximation to the full early time evolution
of any perturbed lattice, treated as a full discrete
$N$-body system. It therefore includes {\it all} modifications 
of the theoretical fluid evolution in its regime of validity,
which extends up to the time when particles approach one
another (i.e. up to close to ``shell crossing''). We will
report elsewhere detailed comparisons in numerical
simulations of this approximation with the ZA and its 
improvements.
In the rest of this letter we consider the 
quantification of the discreteness corrections 
to the pure fluid limit described by our approximation.

Assuming still an EdS universe, and  that the initial perturbations 
are set up in the  standard manner using the ZA, as described above,  
it follows directly from Eq.~(\ref{linearised-evolution-general}) 
that  ${\tilde u}_\mu ({\bf k},t)= \sum_{\nu} {\cal A}_{\mu \nu } ({\bf k},t)
{\tilde u}_\nu ({\bf k},t_0)$, where 
${\cal A}_{\mu \nu} ({\bf k},t)=\sum_{n}
[U_n(t) + \frac{2}{3t_0} V_n(t)]
({\bf{\hat e}}_n)_\mu ({\bf{\hat e}}_n)_\nu $ (the ${\bf k}$ 
dependences on the right hand side are implicit). 
The full linearised evolution is encoded in this matrix,
which can be calculated straighforwardly for any given 
lattice once the eigenvalues and eigenvectors have been
found. One can then determine directly e.g. 
the power spectrum (PS) of the displacement 
fields ${\cal S}_{\mu \nu} ({\bf k}, t)
\equiv {{\tilde u}_\mu}({\bf k},t){{\tilde u}_\nu^*}({\bf k},t)$.
Given ${\cal S}$ one can then calculate, by the method developed 
in \cite{gabrielli-pre-2004}, the PS of the density field 
for the full point distribution. For small displacements
(compared to $l$) and neglecting the terms describing the
discreteness of the lattice, it is a good approximation to use
the continuity equation which gives 
$\tilde {\delta \rho} ({\bf k},t) \approx -i {\bf k}\cdot {\tilde {\bf u}} ({\bf k},t)$, where $\tilde {\delta \rho} ({\bf k},t)$ is the FT of the 
density fluctuation field. It follows that 
$P({\bf k}, t) \approx A_P^2({\bf k}, t) P({\bf k}, t_0)$
where 
$A_P({\bf k}, t)= 
\sum_{\mu, \nu} {\hat k}_\mu {\hat k}_\nu {\cal A}_{\mu \nu}({\bf k},t)$
and $P({\bf k}, t) \propto |\tilde {\delta \rho} ({\bf k},t)|^2$ is the
PS of the density fluctuations. It is simple to verify
that in the fluid limit discussed above ($kl \rightarrow 0$) 
one obtains, as expected, $A_P^2({\bf k},t)= a^2(t)$.

\begin{figure}
\resizebox{8cm}{!}{\includegraphics*{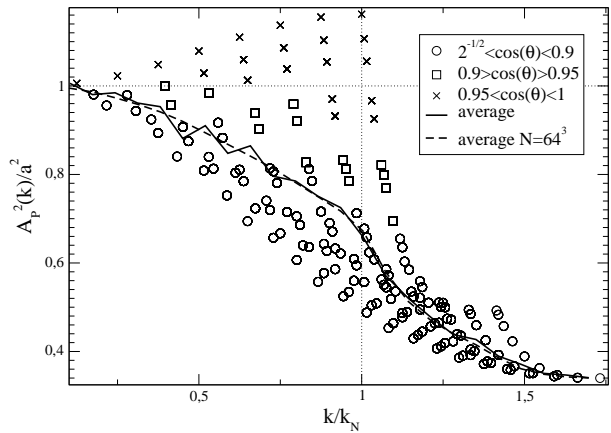}}
\caption{Amplification function $A_P^2(\bk,t)$ for the 
power spectrum, divided by the fluid limit amplification ($a^2$),
at $a=5$, for a sc lattice. See text for details.  
\label{fig2}}
\end{figure}

In Fig. \ref{fig2} is shown this amplification factor 
$A_P^2 ({\bf k},t)$, divided by $a^2$. The scale factor 
chosen is $a=5$, a value at which typical NBS 
reach shell crossing. Deviations from unity are a 
direct measure of the modification of the theoretical
evolution introduced by the discretisation. 
Note that  $A_P^2 ({\bf k},a)$ is plotted as a function of $k$,
each point corresponding to a different value 
of ${\bf k}$.  The fact that the evolution
depends on the orientation of the vector ${\bf k}$ is 
a manifestation of the breaking of rotational invariance 
by the lattice discretisation. The three different symbols
for the points correspond to three different intervals of
the cosine of the minimum angle $\theta$ between 
the vector ${\bf k}$ and one of the axes of the lattice. 
We thus see that the largest eigenvalues correspond to 
modes describing motion parallel to one of the axes
of the lattice. For a $N^3$ lattice and $N$ even,
for instance, the largest eigenvalue, with a growth law 
$\propto a^{1.06}$, is a longitudinal mode with
$k=k_N$ and ${\bf k}$ parallel to the axes of the lattice,
which describes the motion of pairs of adjacent infinite planes 
towards one another. Also shown in the figure 
is an average of $A_P^2 ({\bf k},a)$ over 25 bins of 
equal width in $k$, both for the $16^3$ lattice
from which the points have been calculated, and 
for a larger $64^3$ lattice.

We thus see that there are qualitatively two kinds of
effects introduced by the discretisation: (i) an average
slowing down of the growth of the modes relative to
the theoretical fluid evolution, and (ii) a pronounced 
anistropy in $k$-space. There are notably a small 
fraction of modes (approximately 2.5 \%) with growth 
exponents larger than in linear fluid theory (which, 
for sufficiently large $a$, will always dominate
the evolution). We can conclude, however, as 
foreshadowed in the discussion above, that 
this evidently undesirable feature of the  
sc lattice discretisation can be circumvented
by employing a bcc lattice. The known 
stability  of this configuration 
of the Coulomb lattice \cite{wc-ground state} implies 
that the fluid exponent is in this case an upper bound 
for all modes (and that there are no oscillating modes 
for the case of gravity).
Further the bcc crystal is more isotropic (and indeed
more compact \cite{torquato}) than the sc lattice,
and thus we would expect the effects of breaking of
isotropy to be less pronounced. The average slowing down
of the growth of the modes, by an amount which depends
on the time and the dimensionless product $kl$ (at
$a=5$, as seen in Fig. \ref{fig2},  
a 10\% effect at half the Nyquist frequency), on the other hand, 
would be expected to be
a common feature of any discretisation
(e.g. using ``glassy''  configurations \cite{white-leshouches}, or the 
discretisation developed in \cite{mj-dl-bm-initial-conditions}).

This analysis does not, of course, allow one to 
conclude fully about the role of discreteness
effects in the longer time evolution of such simulations.
It should provide, however, a first step to the quantitative
understanding of these effects. One important
conclusion which we can draw  is  the following: 
for a given physical wavelength, the discrepancy between the fluid
and full evolution grows, up to shell crossing, with time.
{\it Thus, for a given physical scale, discreteness effects 
increase when the starting time of the simulation is decreased}.
For the particular case of the sc lattice it is clear that
artificial collapses of infinite planes along 
the axes of the lattice, 
which grow more rapidly than even
long-wavelength fluid modes,
will be increasingly priveleged as earlier starting
times are taken. 
This implies that at least one of the conditions 
for keeping discreteness effects under control in
an NBS will be a constraint on the
initial time (i.e. that the starting red-shift be 
greater than some value, given the discretisation scale). 
We note that the initial red-shift is not a 
parameter considered in discussions
of discreteness effects in NBS  in the literature 
(e.g. \cite{discreteness-references, discreteness-hamana}).

We can extend our treatment easily to incorporate
a smoothing of the gravitational force up to a scale 
$\epsilon$. Here we have taken pure gravity 
(i.e $\epsilon=0$) as in most cosmological NBS 
$\epsilon \ll l$, which gives negligible modification 
of our results. Just as in the analogous condensed matter 
system, the method can also be  extended to higher order. 
It would be interesting in particular to map at higher order 
this description of the discrete system onto the 
corresponding order of fluid Lagrangian theory, which has 
been explored extensively in the cosmological literature 
(see e.g. \cite{higher-order-lagragian stuff}, and
references therein). Further it should be possible to use the 
approach presented here to understand better the nature of existing
approximations which go beyond the simple fluid limit, for 
example those involving pressure terms associated to velocity dispersion
(see e.g. \cite{higher-order-lagragian stuff, buchert-dominguez}
and references therein).

\end{document}